# Aberration corrected ultraviolet echelle spectrographs: straw man designs and performance

James C. Green
University of Colorado Boulder, Boulder, CO, USA

## ABSTRACT

The Far Ultraviolet (FUV: hereafter 900-1150 Å) is a spectral range which contains many of the ground state transitions of common elements but has had limited observational capabilities due to the unique technological requirements to operate in this waveband. Conceptual designs are presented, for high resolution (R > 50,000) echelle spectrographs for CubeSat, SMEX and MIDEX missions, along with comparisons of their performance to past instruments.

## 1. INTRODUCTION

The most recent decadal survey[1] has prescribed the flagship and probe missions over the next 20 years, and while the IR-optical-UV flagship is planning on spectroscopic capability down to 1000Å, efficient operations below the $MgF_2$ cutoff at ~ 1150Å will require technological improvement in UV reflective coatings. The 900-1000Å regime will remain uncovered in even the most optimistic estimates. The two probe missions will be sub-mm/IR and high energy. There is no observational capability in the FUV regime currently, except the "super-blue" modes of the Cosmic Origins Spectrograph[2] (COS) which has limited capabilities. In a recent paper[3], a solution for an aberration corrected echelle spectrograph were presented. This solution utilized a concave echelle grating ruled through controlled ion-etching[4] and operating in the post-focus diverging beam from a telescope followed by a nearly flat, uniform line-spaced cross disperser with a small amount of curvature to maximize the spectral resolution across the band. Depending on the amount of wavelength coverage desired, and the size of the assumed detector, the system could operate between $\lambda/\Delta\lambda$ = 50,000 to 100,000. This configuration removes one reflective surface from the usual echelle configuration, which is significant given the modest reflectivity available in the FUV, particularly below the LiF cutoff, where silicon carbide (SiC), which has a typical reflectivity of 30% from 900-1000 Å, is the best available option. A fold mirror can be added to increase the throw lengths and dispersion to increase the resolution. While the addition of a folding mirror to increase path lengths can be considered when the available space is limited, this does not mean that a traditional collimator/echelle/cross disperser would then be equivalent. Obtaining the desired dispersion at the detector while controlling the aberrations within the available space would have to be studied in both scenarios to determine the best option.

The fundamental objective of a high-resolution spectrograph operating down to 912Å is to provide sufficient resolution to reduce confusion as the number of strong line in the FUV is large, and to allow profile fitting of absorption and emission lines to better characterize the physics in the gas being studied.

There are some small FUV missions that are approved (e.g., Aspera[5]) but no high resolution FUV spectroscopy mission has been approved. Straw man concepts for CubeSat, SMEX an MIDEX class mission are presented.

## 2. PREVIOUS FUV SPECTROSCOPIC CAPABILITES

### 2.1 Orbiting Astronomical Observatory (OAO-3) *Copernicus*

Copernicus launched in August 1972 and had an ultraviolet spectrometer that operated from 710 – 3275 Å.[6] The system utilized a scanning spectrometer, measuring one wavelength bin at a time which limited its sensitivity. The Al/LiF coating on the optics meant that performance below 1000 Å was significantly less than for wavelengths greater than 1000. The spectral resolution was $\Delta\lambda = 0.06$Å in the UV.

### 2.2 The Hopkins Ultraviolet Telescope (HUT)

HUT launched in 1990 and 1995 on the ASTRO-I an ASTRO II space shuttle missions. The improved HUT II which flew on ASTRO-II covered 415-1850 Å and had a peak effective area of 24 $cm^2$ and resolution $\Delta\lambda$ = 2-4 Å.[7]

### 2.3 The Orbiting and Retrievable Far and Extreme Ultraviolet Spectrograph (ORFEUS-SPAS)

ORFEUS-SPAS was a free-flying, pointed instrument platform that had three separate UV instruments and flew on the space shuttle in 1993 and 1996, each mission lasting ~7 days. The Berkeley EUV/FUV spectrometer operated from 390-1219Å and had a peak effective area of ~9cm and a resolution of ~ 3000.[8] The Interstellar Medium Absorption Profile Spectrograph (IMAPS) was an objective echelle spectrograph that has high resolution ($\Delta\lambda = 0.04$Å) and covered 950-1150 Å with limited sensitivity below 1000Å.[9] The ORFEUS Echelle spectrograph covered 900-1400Å, had a resolution of 10,000 and a peak effective area of 1.9cm$^2$.[10]

### 2.4 The Far Ultraviolet Spectroscopic Explorer (FUSE)

FUSE[11,12] was a MIDEX class mission that launched in 1999 and provided resolution of ~16,000 in two channels. The LiF channel (~1000-1187Å) had 45 cm$^2$ of effective area and the SiC (905 - ~ 1040Å) channel had ~15 cm$^2$. The high sensitivity of FUSE meant that it bright limit was fainter than the faint limits of *Copernicus* and IMAPS, and that no targets could be observed with both systems.

In summary, only Copernicus and FUSE were long duration orbital missions. While HUT and ORFEUS had excellent performance, their observing opportunities were limited to a few days, rather than years, meaning that many science opportunities remain unaddressed. While IMAPS had high spectral resolution, its performance below 1000Å meant this region on the spectrum (900-1000Å) has almost no observations performed. For this reason, the straw man designs below focus on he 900-1000Å region, where even a small mission could provide unprecedented observation capability.

Table 1. Summary of past FUV instrument performance

| Instrument | Sensitivity/Resolution 900-1000Å | Sensitivity /Resolution 1000 – 1200Å |
|---|---|---|
| *Copernicus* | Limited $\Delta\lambda = 0.6$Å | No objects fainter than ~ V=6 $\Delta\lambda = 0.6$Å |
| HUT | 24 cm$^2$ $\Delta\lambda = 2$-4Å | 24 cm$^2$ $\Delta\lambda = 2$-4Å |
| IMAPS | Limited $\Delta\lambda = 0.04$ | Bright stars $\Delta\lambda = 0.04$ |
| ORFEUS/Berkeley | 9 cm$^2$ R ~ 3000 | 9 cm$^2$ R ~ 3000 |
| ORFEUS/Echelle | 0.9 cm$^2$ R ~ 10,000 | 1.9 cm$^2$ R ~ 10,000 |
| FUSE | 15 cm$^2$ R ~ 16,000 | 45 cm$^2$ R ~ 16,000 |

## 3. CONCEPT DESIGNS

These designs focus on maximizing their performance in the 900-1000Å region, as this is the least well served spectral regime for high resolution spectroscopy. Therefore, all surfaces are assumed to have SiC as their optical coating with a reflectivity of 30%. All gratings are assumed to be blazed with 60% groove efficiency (18% absolute efficiency). All detectors are assumed to be microchannel plate systems with a detection quantum efficiency (DQE) of 50% and 25 μm resolution (not pixel size).

## 3.1 CubeSat

While an objective echelle design (a flat echelle grating directly observing the sky) might seem appropriate for a CubeSat to minimize the reflective losses, the requirements for maintaining strict pointing stability are most likely beyond the capabilities of current CubeSat systems. While these will undoubtedly improve with time, the strawman design assumes 12U CubeSat, a prime focus mirror (4 cm square, 16 cm$^2$), and an f/8 beam. The detector is assumed to be a 25mm square active area. This design will yield 15,000 resolution across the entire 900-1150 band simultaneously and provides ~0.08 cm$^2$ of effective area. While this number is small, the integration times can be large, and the potential targets (hot, nearby stars) are bright. For example, ε Canis Majoris, a nearby B2-II star with V = 1.5 would produce about 10 counts/second/resolution element, resulting in a high signal-to-noise spectrum in a short time. With the addition of an additional fold mirror, the throw lengths and corresponding resolution can be increased to 50,000. Even with reducing the effective area by a factor of 0.3 and reducing the spectral width by a factor of three, the instrument would still produce a signal of ~1 count/second/resolution element, or a S/N = 100 spectrum (assuming signal limited) in about 3 hours.

## 3.2 SMEX

A SMEX concept assumes that a 0.5m diameter mirror and a post telescope focus length of 1.8m is available for the spectrograph. The detector is assumed to be 50mm square. The telescope is assumed to be two-mirror, so that the f/# can be chosen by design. The system is designed to operate at 0.8Å/mm which is 0.02Å/resolution element or R=50,000. The effective area is 2.4cm$^2$ across the 900-1000Å regime. While less sensitive than HUT or ORFEUS/Berkeley, it has significantly higher spectral resolution. As an orbital mission, it would have significant total observing time advantages over HUT and ORFEUS and could observe more and fainter targets. However, this improvement may not be sufficient to justify a dedicated mission.

## 3.3 MIDEX

A MIDEX concept assumes a 1m diameter optic, and 2.5m of space available behind the telescope focus. The larger space allows longer throw lengths and more dispersion, allowing 0.4Å/mm and a 0.01Å resolution element (R = 100,000 at λ = 1000Å) It is essentially a scaled up MIDEX, and gains from the larger telescope, giving it ~10cm$^2$ of effective area. This instrument would allow a new class of observations, with high S/N, high resolution studies of extragalactic objects and faint/confused/overlapping absorption features in the ISM, CGM and IGM.

## 4. CONCLUSIONS

The spectrograph concepts outlined here assume that all optical surfaces are coated with SiC, including the telescope. Unless new coatings are developed that provide excellent reflectivity in the Optical/UV as well as the FUV, the addition of a SiC based spectrograph to a mission focused on the UV and/or optical will not be feasible. A spectrograph optimized for the 900-1000Å band could be considered as a whole telescope/spectrograph add-on or a stand-alone mission. Given the limited number of stand-alone mission opportunities, the author considers it unlikely that a SMEX or MIDEX mission could go forward. However, utilizing these concepts, a CubeSat or Pioneer mission could provide unique and valuable high-resolution data or nearby hot stars and the local interstellar medium that can not be obtained from the ground or sub-orbital missions.